\documentclass[twocolumn,prl,aps,longbibliography,superscriptaddress,floatfix,tightenlines]{revtex4-2}
\usepackage{amsmath}
\usepackage{amssymb}
\usepackage{mathrsfs}
\usepackage{amsfonts}
\usepackage{epsfig}
\usepackage{bm}
\usepackage{times}
\usepackage{lineno}
\usepackage[T1]{fontenc}
\usepackage[dvips]{color}
\usepackage[colorlinks,bookmarks=false,citecolor=blue,linkcolor=magenta,urlcolor=blue]{hyperref}
\begin{document}

\title{Multipartite Entanglement of the Topologically Ordered State in a Perturbed Toric Code}
\author{Yu-Ran Zhang}
\affiliation{Theoretical Quantum Physics Laboratory, RIKEN Cluster for Pioneering Research,
Wako-shi, Saitama 351-0198, Japan}

\author{Yu Zeng}
\affiliation{Institute of Physics, Chinese Academy of Sciences, Beijing 100190, China}

\author{Tao Liu}
\affiliation{School of Physics and Optoelectronics, South China University of Technology, Guangzhou 510640, China}
\affiliation{Theoretical Quantum Physics Laboratory, RIKEN Cluster for Pioneering Research,
Wako-shi, Saitama 351-0198, Japan}

\author{Heng Fan}
\email{hfan@iphy.ac.cn}
\affiliation{Institute of Physics, Chinese Academy of Sciences, Beijing 100190, China}
\affiliation{CAS Center for Excellence in Topological Quantum Computation, UCAS, Beijing 100190, China}

\author{J. Q. You}
\email{jqyou@zju.edu.cn}
\affiliation{Department of Physics, Zhejiang University, Hangzhou 310027, China}

\author{Franco Nori}
\email{fnori@riken.jp}
\affiliation{Theoretical Quantum Physics Laboratory, RIKEN Cluster for Pioneering Research, Wako-shi,
Saitama 351-0198, Japan}
\affiliation{RIKEN Center for Quantum Computing (RQC),  Wako-shi, Saitama 351-0198, Japan}
\affiliation{Physics Department, University of Michigan, Ann Arbor, Michigan 48109-1040, USA}


\begin{abstract}
We demonstrate that multipartite entanglement, witnessed by the quantum Fisher information (QFI), can characterize
topological quantum  phase transitions in the spin-$\frac{1}{2}$ toric code model on a square lattice with
external fields. We show that the QFI density of the ground state can be written in
terms of the expectation values of gauge-invariant Wilson loops for different sizes of square regions and identify
$\mathbb{Z}_2$ topological order by its scaling behavior. Furthermore, we use this multipartite entanglement witness
to investigate thermalization and disorder-assisted stabilization of topological order after a quantum
quench. Moreover, with an upper bound of the QFI, we demonstrate the absence of finite-temperature
topological order in the 2D toric code model in the thermodynamic limit.
Our results provide insights to 
topological phases, which are robust against external
disturbances, and are candidates for topologically protected  quantum computation.

\end{abstract}

\maketitle

 \emph{Introduction.---}Recent  developments of quantum information theory have provided novel angles and tools
for  modern condensed matter physics \cite{Zeng2015}. The quantum-information features of
quantum matter   help to study quantum phase transitions (QPTs) \cite{Vojta2003,Wen2004,Amico2008,Wen2017,Liu2016,Zhu2019},
out-of-equilibrium quantum many-body systems \cite{Eisert2010,Eisert2015,Gogolin2016}, and non-Hermitian physics
\cite{Lee2014,Feng2017,ElGanainy2018,Oezdemir2019,Matsumoto2020,Ashida2020}.
One pivotal concept introduced by quantum information science into  condensed matter physics
is quantum entanglement \cite{Horodecki2009}, which quantifies 
\emph{nonlocal} quantum correlations.
Beyond Landau's symmetry-breaking theory \cite{Wen2004}, topological order \cite{Wen1990} in
strongly correlated many-body systems, e.g., quantum spin liquids \cite{Zhou2017} and the
fractional quantum Hall states \cite{Stormer1999}, can be described by the long-range entanglement
encoded in the ground states of the system \cite{Zeng2015}. Topological order is promising
for applications in fault-tolerant quantum computation \cite{Kitaev2003,Nayak2008},
because of properties such as ground-state degeneracy and the non-Abelian geometric phase of degenerate
ground states.
%
As a simple example of a stabilizer code, the toric
code model (TCM), with fourfold topological degeneracy on a torus, is the widest studied model of topological order,
allowing for encoding two robust qubits  against \emph{local} perturbations \cite{Kitaev2003}.

Bipartite entanglement witnesses, including  topological entanglement entropy
\cite{Kitaev2006,Levin2006}, entanglement spectrum \cite{Li2008}, topological R\'{e}nyi entropy
\cite{Flammia2009,Halasz2013}, and entanglement negativity \cite{Castelnovo2013,Lee2013}, have
achieved a great success in characterizing  topological order.
However, the measurement of bipartite entanglement requires  quantum
state tomography or a statistical protocol via randomized measurements \cite{Brydges2019},
which are almost impossible even for a modest system size.
Multipartite entanglement, witnessed by the quantum Fisher information (QFI)
\cite{BRAUNSTEIN1994,Ma2011,Guehne2009}, is
believed to represent richer properties of complex structures of topological states than bipartite entanglement,
which is experimentally measurable in large many-body systems using  mature
techniques \cite{Hauke2016,Strobel2014,Pezze2018}.
Recent studies \cite{Pezze2017,Zhang2018}
have shown that the scaling behavior of the QFI with respect to \emph{nonlocal} operators is sensitive when detecting
1D symmetry-protected topological (SPT) order \cite{Gu2009,Chen2011} and topological order in the
Kitaev honeycomb model \cite{Kitaev2006a}.

In this Letter,
we demonstrate the characterization of topological QPTs in a 2D spin-$\frac{1}{2}$ TCM with external
fields by using  multipartite entanglement, witnessed by the QFI. With a dual
transformation \cite{Zhang2018}, the QFI density of the ground state can be expressed in terms of the
reduced Wilson loops \cite{Halasz2012} for different sizes of square regions, whose scaling behavior signals the
topological QPTs. Furthermore,  we show that thermalization and  disorder-assisted stabilization of topological
order after a quantum quench can be identified via  multipartite entanglement. Using an upper bound of the QFI, we investigate the 2D TCM at finite temperatures and show that topological order cannot survive against
thermal fluctuations in the thermodynamic limit. Our work,
based on the experimentally extractable QFI, will contribute to a deeper understanding of topological QPTs in condensed matter physics and has promising applications in both fault-tolerant quantum computation and robust quantum metrology \cite{Giovannetti2011}.

\begin{figure}[t]
\centering
\includegraphics[width=0.49\textwidth]{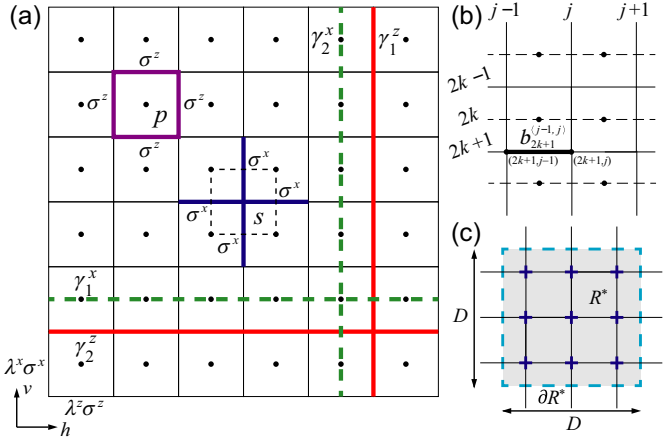}\\
\caption{(a) A $N \times N$ square lattice with periodic boundary conditions and
spins-$\frac{1}{2}$ on the bonds. A star ($s$) on the lattice corresponds to a plaquette
($p$) on the dual lattice and vice versa. Fields in the $z$ $(x)$ direction with a magnitude of
$\lambda^{z}$ ($\lambda^x$) are located on the horizontal (vertical) edges. (b)
An illustration of sites $(i,j)$, where the effective spins are located, and links
$b^{\langle j-1,j\rangle}_{i}$. (c) An illustration of the magnetic
Wilson loop for a square region $R^*$, with $D=3$. The region $R^*$ contains $D\times D$ stars
(dark blue crosses), and the boundary $\partial R^*$ (light blue dashed line) threads $4 D$ spins.
}\label{fig:1}
\end{figure}

\emph{The TCM on a torus with external fields.---}%
The TCM has a resonating-valence-bond phase \cite{ANDERSON1196} (a.k.a. a quantum spin liquid
phase \cite{Wen2004,Zeng2015}), capturing all elements of topological order: $\mathbb{Z}_2$ excitations with
anyonic particle statistics, a fourfold degenerate ground state on a torus, robustness against \emph{local} perturbations,
and no \emph{local} order parameter.
Similar to the quantum dimer model on the Kagome lattice \cite{Misguich2002} and the
Wen-plaquette model \cite{Wen2003}, the 2D TCM is a significant toy model for studying
$\mathbb{Z}_2$ topological order.
The Hamiltonian of the spin-$\frac{1}{2}$ TCM on a $N\times N$ square lattice, reads \cite{Kitaev2003,Zeng2016}
\begin{equation}
\hat{H}_{\textrm{tc}}=-J^A\sum_s\hat{A}_s-J^B\sum_p\hat{B}_p,\label{ham}
\end{equation}
where the stabilizer operators $\hat{A}_s\equiv\prod_{i\ni s}\hat{\sigma}_i^x$ and
$\hat{B}_p\equiv\prod_{i\in p}\hat{\sigma}_i^z$ both contain four spins belonging to a star $s$ and a
plaquette $p$, respectively [see Fig.~\ref{fig:1}(a)].
This model is exactly solvable, because all
the star and plaquette operators commute with each other, i.e., $[\hat{A}_s,\hat{B}_p]=0$ for $\forall s,p$.
We have two constraints, $\prod_s\hat{A}_s=\prod_p\hat{B}_p={\mathbb{I}}$,
with periodic boundary conditions, and thus, the ground state is fourfold degenerate, allowing for
encoding  two qubits.
Any state in the ground-state manifold $\mathscr{L}$ can be written as $|\mathcal{G}\rangle=\sum_{i,j=0}^1a_{ij}|\xi_{ij}\rangle$
in terms of four bases $|\xi_{ij}\rangle=(\hat{W}_1^{\textrm{m}})^i(\hat{W}_2^{\textrm{m}})^j|\xi_{00}\rangle$, where we have $\sum_{i,j=0}^1|a_{ij}|^2=1$, the string operators $\hat{W}_{1,2}^{\textrm{m}}\equiv\prod_{i\in\gamma_{1,2}^x}\hat{\sigma}_i^x$, and
$|\xi_{00}\rangle\varpropto\prod_{s}({{\mathbb{I}}+\hat{A}_{s}})\!\!\mid\Uparrow\rangle$,
with $\mid\Uparrow\rangle$ being the all spin-up state in the $\hat{\sigma}^z$ basis.
The excitations of the TCM are in two categories:
the electric charges (e) and magnetic vortices (m) of a $\mathbb{Z}_2$ lattice gauge theory, which have
non-trivial mutual statistics and follow fusion rules  \cite{Kitaev2003}.

Then we consider the system (\ref{ham}) subsequently subject to the  fields in the $z$ and $x$ directions on the
horizontal and vertical edges, respectively [see Fig.~\ref{fig:1}(a)]:
\begin{equation}
\hat{V}=-\sum_{i\in h}\lambda^z_i\hat{\sigma}_i^z-\sum_{j\in v}\lambda^x_j\hat{\sigma}_j^x.\label{field}
\end{equation}
The Hamiltonian of the TCM with external fields can be expressed
as uncoupled Ising chains on different lines: \cite{Yu2008,Zhang2011}
\begin{align}
\hat{H}_{\textrm{tc}}^{\textrm{field}}=\hat{H}_{\textrm{tc}}+\hat{V}=\sum_{i=1}^{2N}\hat{{H}}_{i},\label{ptcm}
\end{align}
where the transverse-field Ising Hamiltonian
is $\hat{{H}}_{i}\equiv-\sum_{j=1}^N[J^{(i)}\hat{\tau}^x_{i,j}
\hat{\tau}^x_{i,j+1}
+{\lambda}_j^{(i)}\hat{\tau}^z_{i,j}
]$,
with
 ${\lambda}^{(i)}_j=\lambda_j^{z}$ ($\lambda_j^{x}$) and $J^{(i)}=J^A$ ($J^B$) for $i$ being odd (even),
$[\hat{{H}}_{i},\hat{{H}}_{l}]=0$, and $\hat{\tau}^{x,z}_{i,j}$
being Pauli operators, after
applying the dual transformation \cite{SM}, on
the site $(i,j)$ 
of the original and dual lattice at the $i$th row  and $j$th column  [see Fig.~\ref{fig:1}(b) for the original lattice].


\emph{Gauge-invariant Wilson loop.---}%
A $\mathbb{Z}_2$ lattice gauge theory \cite{Kogut1979} has confined and deconfined phases,
and is equivalent to the classical Ising model.
The TCM \cite{Halasz2012} 
can be mapped into the
Hamiltonian of a $\mathbb{Z}_2$ lattice gauge theory \cite{Fradkin2013}, and the $\mathbb{Z}_2$ topological order (in the deconfined phase)
can be probed via the expectation value of the gauge-invariant  Wilson loop \cite{Wegner1971,Kogut1979}.
For a square region $R^*$  on the dual lattice with $D\times D$ stars [see Fig.~\ref{fig:1}(c)], the Wilson loop is
defined as the average of a magnetic closed string operator $\mathcal{W}^{\textrm{m}}_{R^*}\equiv\langle\hat{{W}}^{\textrm{m}}_{R^*}\rangle$ \cite{Halasz2012}, with
\begin{equation}
\hat{{W}}_{R^*}^{\textrm{m}}=\prod_{i\in\partial R^*}\hat{\sigma}_i^x=\prod_{s\in R^*}\hat{A}_s,
\end{equation}
where $\partial R^*$ denotes the boundary of $R^*$. Similarly, the Wilson loop for a square region $R$ on the original  lattice,
with $D\times D$ plaquettes, can be defined as $\mathcal{W}_R^{\textrm{e}}=\langle\hat{{W}}_R^{\textrm{e}}\rangle$,
with an electric closed string operator $\hat{{W}}_{R}^{\textrm{e}}=\prod_{i\in\partial R}\hat{\sigma}_i^z=\prod_{p\in R}\hat{B}_p$.
For simplicity, we consider the reduced Wilson loop, defined as
$w_D^{\textrm{e,m}}\equiv(\mathcal{W}^{\textrm{e,m}}_D)^{{1}/{D}}$.
For a large square region with $D\gg 1$, the Wilson loops on both original and dual lattices follow a
  perimeter law
  $\mathcal{W}^{\textrm{e,m}}\propto\exp(-\beta D)$, i.e., $w_D^{\textrm{e,m}}\rightarrow \textrm{const}\neq0$,  for the existence of topological order; and either or both
  follow an area law $\mathcal{W}^{\textrm{e}}$ or $\mathcal{W}^{\textrm{m}}\propto\exp(-\beta D^2)$, i.e., $w_D^{\textrm{e}}$ or $w_D^{\textrm{m}}\rightarrow0$,  when topological order is absent \cite{Halasz2012}.

\begin{figure}[t]
	\centering
	\includegraphics[width=0.49\textwidth]{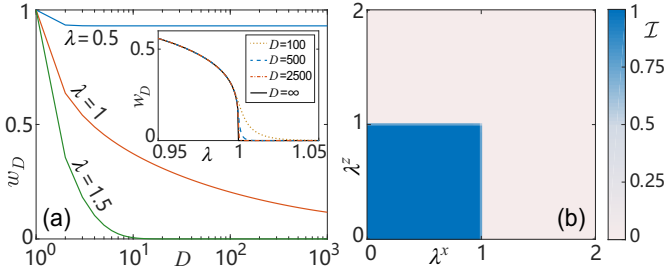}\\
	\caption{
		(a) The reduced Wilson loop $w_D$ as a function of the side length $D$ of the square region
		   on the  lattice of the TCM
		for different  strengths, $\lambda=0.5,1,1.5$, of the external fields.
		(b) The scaling topological index  $\mathcal{I}\equiv \beta^{\textrm{e}}\beta^{\textrm{m}}$ obtained from
		 the QFI densities
		$f_Q[\mathcal{O}^{\textrm{e,m}},|\mathcal{G}\rangle]$ of a chosen square region with a side length $L=\textrm{2,000}$
		for different strengths $\lambda^{x,z}$
		of external fields. The size of the square lattice of the TCM is $N\times N$ with $N=5L=\textrm{10,000}$.
	}\label{fig:2}
\end{figure}

For the TCM with external fields, the reduced Wilson loop can be written as a spin-spin correlator  for the ground state,
\begin{equation}
w_D^{{\textrm{e,m}}}=\langle\hat{\tau}^x_{i,j}
\hat{\tau}^x_{i,j+D}
\rangle_\mathcal{G},\label{reduced}
\end{equation}
for $i$ being even and odd, respectively. 
For simplicity, we consider uniform  external fields $\lambda_j^{x,z}=\lambda^{x,z}$ and
set $J_A=J_B=1$.
For the topologically trivial phase  ($\lambda^{x}$ or $\lambda^{z}>1$), we have $w_D^{\textrm{e}}$ or $w_D^{\textrm{m}}\rightarrow 0$; for
the topologically non-trivial phase with topological order   ($\lambda^{x,z}<1$),
we have $w_D^{\textrm{e,m}}\rightarrow \textrm{const}>0$ [see Fig.~\ref{fig:2}(a)].



\emph{Probing topological order in the TCM by the  QFI density.---}%
Recent studies \cite{Pezze2017,Zhang2018} show that multipartite entanglement, witnessed by the QFI
with \emph{nonlocal} operators, can characterize 1D SPT order, and topological order in the Kitaev honeycomb model.
For a pure state $|\psi\rangle$, the QFI, with respect to a generator $\hat{\mathcal{O}}$, can be simply calculated as
$F_Q=4(\Delta_\psi \hat{\mathcal{O}})^2$ \cite{BRAUNSTEIN1994}, where
$(\Delta_\psi \hat{\mathcal{O}})^2\equiv\langle\hat{\mathcal{O}}^2\rangle_\psi-\langle\hat{\mathcal{O}}\rangle_\psi^2$.
For an $m$-partite system, the QFI density, defined as $f_Q\equiv F_Q/m$, gives  an  entanglement criterion:
 the violation of the inequality $f_Q \leq\kappa$ signals ($\kappa+1$)-partite entanglement
 ($1 \leq \kappa \leq m - 1$) \cite{Pezze2009}.

For the TCM with uniform external fields in the $x, z$ directions, we consider the square
regions with $L\times L$ spins on the original and dual lattices, respectively.  The generators
for the QFI are chosen as
\begin{equation}
\hat{\mathcal{O}}^{\textrm{e}}=\sum_{k,j=1}^L
\hat{\tau}^{x}_{2k,j}/2,
\hspace*{0.2 in}\hat{\mathcal{O}}^{\textrm{m}}=\sum_{k,j=1}^L
\hat{\tau}^{x}_{2k-1,j}/2,
\label{generator}
\end{equation}
for the regions on the original and dual lattices, respectively.
We obtain two QFI densities:
\begin{equation}
f_Q[\hat{\mathcal{O}}^{\textrm{e,m}},|\mathcal{G}\rangle]\equiv F_Q[\hat{\mathcal{O}}^{\textrm{e,m}},|\mathcal{G}\rangle]/L^2=1+\sum_{D=1}^{L-1}w_D^{\textrm{e,m}},
\end{equation}
which are expressed in terms of the reduced Wilson loops (\ref{reduced}) for increasing sizes  of the square regions on the original and dual lattices, respectively.
As discussed in Refs.~\cite{Hauke2016,Pezze2017,Zhang2018}, the QFI densities, as a function of $L$,
follow an asymptotic power-law scaling in the thermodynamic limit:
\begin{equation}\label{scaling}
f_Q[\hat{\mathcal{O}}^{\textrm{e,m}},|\mathcal{G}\rangle]=1+\alpha^{\textrm{e,m}}L^{\beta^{\textrm{e,m}}},
\end{equation}
where the scaling coefficients $\alpha^{\textrm{e,m}}$ and $\beta^{\textrm{e,m}}$ depend on the
parameters of the TCM with external fields. When $\beta^{\textrm{e,m}}\rightarrow1$, multipartite
 entanglement, witnessed via the QFI densities, increase linearly with the side length $L$ of the region,  indicating the presence of long-range quantum correlations.

Then, we can define a topological index as $\mathcal{I}\equiv \beta^{\textrm{e}}\times\beta^{\textrm{m}}$,
combining the scaling behaviors of the QFI densities with respect to the generators (\ref{generator})
for the original and dual lattices, respectively. 
According to the scaling behavior of the Wilson loop,
this topological index approximates 1
in the topological phase 
($0<\lambda^{x,z}<1$), and vanishes in the
topologically trivial phase ($\lambda^{x}>1$ or $\lambda^{z}>1$). Here, the absence of
\emph{nonlocal} entanglement in the regions  on the original lattice $\beta^{\textrm{e}}\simeq0$ (dual lattice $\beta^{\textrm{m}}\simeq0$) indicates
the existence of many anyonic electric (magnetic) excitations, characterized by the Wilson loop
according to the $\mathbb{Z}_2$ lattice gauge theory. Thus,  $\mathcal{I}\rightarrow1$, signaling
multipartite entanglement with respect to the generators defined in both original and dual lattices, indicates the existence of  topological order in the TCM with external fields, and $\mathcal{I}\rightarrow0$ implies
the absence of topological order.
As shown in Fig.~\ref{fig:2}(b), the topological QPTs in the
2D TCM with external fields can be effectively characterized via  multipartite entanglement.

\emph{Observing thermalization of the topological state after a quantum quench via  multipartite entanglement.---}%
Topological order, promising for topological quantum memories \cite{Dennis2002}
and topological quantum computation
\cite{Nayak2008}, is believed to be robust against perturbations,
which cannot change the topological nature of the ground state.
However,  topological order in the 2D TCM is not stable when kicked out of equilibrium \cite{Kay2009,Zeng2016} or at finite temperatures \cite{Brown2016}.
First, we use multipartite entanglement to study thermalization of the topological
ground state of the TCM, after a quantum quench with external fields.
The system before the sudden quench ($t<0$) is supposed to be in the topologically ordered
ground state, $|\Psi_0\rangle=|\mathcal{G}_{\textrm{tc}}\rangle$, of the TCM with $\hat{H}_{\textrm{tc}}$ in Eq.~(\ref{ham}).
At $t=0$, the Hamiltonian is suddenly changed to $\hat{H}_{\textrm{tc}}^{\textrm{field}}=\hat{H}_{\textrm{tc}}+\hat{V}$,
with $\hat{V}$ in Eq.~(\ref{field}) being the uniform external fields in the $x, z$ directions
($\lambda^{x,z}_j=\lambda^{x,z}$). The system evolves as $|\Psi(t)\rangle=\exp(-it\hat{H}_{\textrm{tc}}^{\textrm{field}})|\mathcal{G}_{\textrm{tc}}\rangle$,
and we focus on the stable state of the system at an infinite-long time ($t\rightarrow\infty$).
In the thermodynamic limit $N\rightarrow\infty$,
for a large chosen region of the original (dual) lattice  $D\gg1$,
the long-time  reduced Wilson loop can be calculated as \cite{SM,Suzuki2012}
\begin{equation}
w_D^{\textrm{e,m}}=\left\{\begin{array}{l l}
[({1+\sqrt{1-\lambda^2}})/{2}]^{D},&\textrm{for }~0<\lambda^{x,z}\leq1,\\
~~~~~~~~~~~~1/2^D,&\textrm{for }~\lambda^{x,z}>1,
\end{array}\right.
\end{equation}
which decays to zero for $D\rightarrow \infty$, except for the case without quenched external fields
$\lambda^{x}=\lambda^{z}=0$.
Thus,
for any non-zero quenched external field, $\lambda^{x}$ or $\lambda^{z}\neq0$,
we can derive that the QFI density $f_Q[\hat{\mathcal{O}}^{\textrm{e}},|\Psi(t\rightarrow\infty)\rangle]$
or $f_Q[\hat{\mathcal{O}}^{\textrm{m}},|\Psi(t\rightarrow\infty)\rangle]=\textrm{const}$, leading
to thermalization of topological order 
with a zero topological index
$\mathcal{I}\simeq0$ (see also Supplementary Material \cite{SM}). 
Therefore, thermalization of the topologically ordered state witnessed via
multiparite entanglement implies that the quench of
any transverse field is a source of fluctuations for destroying topological order
in the TCM at zero temperature.

\emph{Dynamical localization of the topologically ordered state with disorder.---}%
We now investigate how to stabilize the topologically ordered state in the TCM using the multipartite entanglement witness.  By introducing disorder in the coupling strengths of stabilizer operators \cite{Wootton2011,Stark2011,Kay2011},
topological order can be protected  from a quantum quench with external fields at zero temperature
\cite{Zeng2021}.
The Hamiltonian of the TCM with disordered coupling strengths reads
\begin{equation}
\hat{H}_{\textrm{tc}}^{\textrm{D/O}}=-\sum_sJ^A_s\hat{A}_s-\sum_pJ^B_p\hat{B}_p,\label{dyham}
\end{equation}
where the random coupling strengths  $J^{A}_{s}=J^A+\delta J_s^A$ and
$J^{B}_p=J^B+\delta J^B_p$ are applied, with $\delta J^A_s\in[-\delta J^{A},\delta J^{A}]$ and
$\delta J^B_p\in[-\delta J^{B},\delta J^{B}]$.
Given $J^{A,B}>0$ and $0<\delta J^{A,B}<1$, the system is stable in the same ground state $|\mathcal{G}_{\textrm{tc}}\rangle$
of the TCM (\ref{ham}) at zero temperature. At time $t=0$, the sudden quench dynamics occurs by
adding external fields $\hat{V}$ (\ref{field})  at $t=0$.
It has been demonstrated in Ref.~\cite{Zeng2021} that in the presence of  disorder in the coupling strengths of stabilizer
operators, the quantum quench is equivalent to a quasi-adiabatic evolution with a family of \emph{local} Hamiltonians,
and the quenched state
$|\Psi(t)\rangle=\exp[-i(\hat{H}_{\textrm{tc}}^{\textrm{D/O}}+\hat{V})t]|\mathcal{G}_{\textrm{tc}}\rangle$
and the initial ground state $|\mathcal{G}_{\textrm{tc}}\rangle$ belong to the same topological phase
\cite{Hastings2005,Osborne2007,Bravyi2010,Chen2010}.
Therefore, after a quantum quench, dynamical localization, induced by disorder, can preserve
the ground state degeneracy,  energy gap, and topological order robustness  \cite{Zeng2021}.

\begin{figure}[t]
	\centering
	\includegraphics[width=0.46\textwidth]{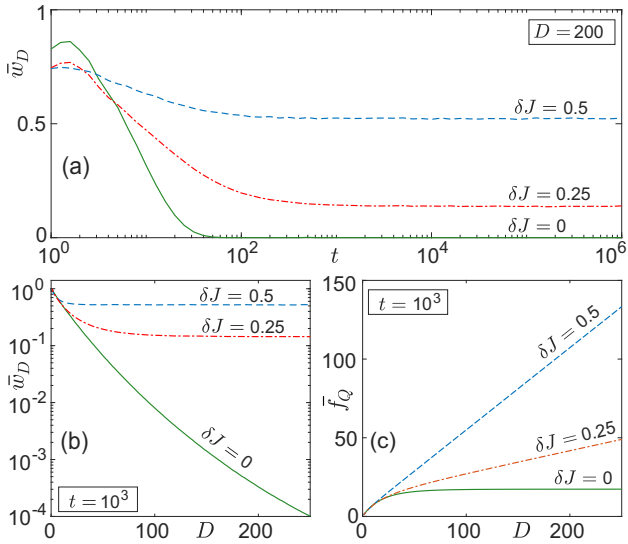}\\
	\caption{Dynamical localization of topological order in the 2D TCM after a quantum quench.
  (a) Time evolutions of the average reduced Wilson loops $\bar{w}_D$ for $\delta J=0,0.25,0.5$,
   with a length $D$ of the square region up to $L=200$.
  (b) The average reduced Wilson loops $\bar{w}_D$ versus the length $D$ of the square region  for $\delta J=0,0.25,0.5$ at $t=10^3$.
  (c) The average QFI densities $\bar{f}_Q$ versus the length $D$ of the square region for  $\delta J=0,0.125,0.25,0.5$ at $t=10^3$.
  The side length of the square lattice is $N=5L=\textrm{1,000}$, and the number of realizations of disordered coupling strengths is 1,000.
	}\label{fig:5}
\end{figure}

We now numerically investigate the dynamical localization of topological order by
considering multipartite entanglement of the
time-evolved state. Without loss of generality, we set $J^{A,B}=1$, consider the same disorder strength
$\delta J^{A,B}=\delta J$, and choose the strengths of the quenched fields in Eq.~(\ref{field}) as
$\lambda_i^z=\lambda_j^x=0.5$.
In Fig.~\ref{fig:5}(a), we plot the time evolutions of the average reduced Wilson loops $\bar{w}_D$ over 1,000
realizations of disorder coupling strengths, with a length of the square region up to $L=200$ and
a length of the square lattice being $N=5L=\textrm{1,000}$.
For long-time evolutions, the average reduced Wilson loops converge to a finite value (with disorder),
or exponentially decay (without disorder).
For long times (e.g., $t=10^3$), the average Wilson loops of the stable states follow a perimeter law (with
disorder) or an area law (without disorder) [see Fig.~\ref{fig:5}(b)].
Thus, the average QFI densities $\bar{f}_Q\propto L$ (with disorder) or $\bar{f}_Q\rightarrow \textrm{const}$  (without disorder)  [see Fig.~\ref{fig:5}(c)].
In addition to the numerical results using topological entanglement entropy \cite{Zeng2021}, we show that
the stable states have the same scaling behaviors of multipartite entanglement as the topological ground states
of the TCM. These results manifest that these two kinds of states belong to the same topological phase,
and  topological order can be protected from a quantum quench by introducing disorder.


\emph{Absence of topological order in the thermodynamic limit at any finite temperature.---}%
Since the topological entanglement entropy cannot distinguish quantum correlations from classical ones
for a mixed thermal state, it has attracted growing interest to search for an order parameter, based on a mixed-state entanglement detection,
for finite-temperature topological order  \cite{Castelnovo2007,Hart2018,Lu2020}.
Here, using  the QFI as an effective entanglement witness for a mixed state \cite{Guehne2009,Lu2020b}, we investigate whether
topological order can survive in the 2D TCM  against thermal fluctuations at finite temperatures.

For a mixed state $\rho$ with all possible pure-state ensembles $\{p_l,|\phi_l\rangle\}$,
the QFI is given by the convex roof of the averaged variance of the generator $\hat{\mathcal{O}}$
\cite{Yu2013}, i.e.,
\begin{equation}
F_Q[\hat{\mathcal{O}},\rho]=4\min_{\{p_l,|\phi_l\rangle\}}\sum_l p_l(\Delta_{\phi_l}\hat{\mathcal{O}})^2.
\end{equation}
To better approximate the QFI of a thermal state of the TCM  at finite temperatures, we use the minimally entangled typical
quantum states (METTSs)
\cite{White2009} to decompose the thermal state  as \cite{Gabbrielli2018,Hauke2016}
$\rho_T=\exp({-\hat{H}_{\textrm{tc}}/T})/\mathcal{Z}=\sum_{i}p_l|\phi_l\rangle\langle\phi_l|$,
 where $\mathcal{Z}\equiv\textrm{Tr}[\exp({-\hat{H}_{\textrm{tc}}/T})]$ and $p_l=\langle l|\rho_T|l\rangle$. Here,
 $|\phi_l\rangle =\exp[-\hat{H}_{\textrm{tc}}/(2T)]|l\rangle/\sqrt{p_l}$
is the METTS by taking $|l\rangle$ as a product state.
The central idea of METTSs is to break a thermal state with inverse temperature $1/T$ into two
copies of the thermal state with $1/(2T)$ \cite{White2009}.
For the square region $R^*$ on the dual lattice, we choose $|l\rangle$
as a product state
in the $\hat{\sigma}^z$ basis and have $|\phi_l\rangle\propto \exp[{ J^A\sum_s\hat{A}_s}/(2T)]|l\rangle$. The QFI density is upper bounded as
$f_Q[\hat{\mathcal{O}}^{\textrm{m}},\rho_T]\leq1+\sum_{D=1}^{L-1} \tanh(J^A/T)^D$, which converges
to a constant for any non-zero temperature $T>0$ in the thermodynamic limit $L\rightarrow\infty$.
Similarly, for the region $R$ on the original lattice, we choose $|l'\rangle$ to be the product state
in the $\hat{\sigma}^x$ axis, and the QFI density is upper bounded as
$f_Q[\hat{\mathcal{O}}^{\textrm{e}},\rho_T]\leq1+\sum_{D=1}^{L-1} \tanh(J^B/T)^D
\rightarrow\textrm{const}$, for
$T>0$ and $L\rightarrow\infty$.
Moreover, with these upper bounds, we can simply deduce that the scaling coefficients of the QFI densities,
defined in Eq.~(\ref{scaling}), are both equal to zero,
${\beta}^{\textrm{e,m}}=0$, with a zero topological index,
${\mathcal{I}}=0$, at any finite temperature.
Therefore, we conclude that multipartite entanglement, witnessed via the QFI, can detect the absence of finite-temperature topological order in the 2D TCM in the thermodynamic limit,
which agrees with the result obtained using negativity \cite{Lu2020}.

%
%
\emph{Conclusions.---}%
We demonstrated that the QFI, as a witness of  multipartite entanglement, can characterize topological QPTs in the
TCM on a square lattice with external fields. The QFI density of the ground state is expressed in
terms of the expectation values of reduced Wilson loops for different sizes of square regions,
of which the scaling behavior identifies the $\mathbb{Z}_2$ topological order. Moreover,  the QFI can be used
to study thermalization and disorder-assisted stabilization of topological order after a quantum
quench. Last, the convex roof of the QFI is applied to
investigate  the TCM at finite temperatures, showing that  topological order
cannot survive at any nonzero temperature in the thermodynamic limit.
Using the experimentally extractable QFI, our results will help to study topological QPTs in condensed matter physics with  promising applications in fault-tolerant quantum computation and robust quantum metrology.

\begin{acknowledgments}
We would like to thank Dr. Zi-Yong Ge for useful discussions.
Y.R.Z. is partially supported by the Japan Society for the Promotion of Science (JSPS)
(via the Postdoctoral Fellowship Grant No.~P19326, and the KAKENHI Grant No.~JP19F19326).
H.F. is partially supported by
the National Natural Science Foundation of China (NSFC) (Grant Nos.~11774406, and 11934018),
and the Strategic Priority Research Program of Chinese Academy of Sciences (Grant No.~XDB28000000).
T.L.  thanks the support from the startup grant of South China University of Technology.
J.Q.Y. is partially supported
by the National Key Research and Development Program of China (Grant No.~2016YFA0301200),
the National Natural Science Foundation of China (NSFC) (Grants Nos.~11934010, and U1801661),
and the Zhejiang Province Program for Science and Technology (Grant No.~2020C01019).
F.N. is supported in part by:
Nippon Telegraph and Telephone Corporation (NTT) Research,
the Japan Science and Technology Agency (JST) [via
the Quantum Leap Flagship Program (Q-LEAP),
the Moonshot R\&D Grant No.~JPMJMS2061, and
the Centers of Research Excellence in Science and Technology (CREST) Grant No.~JPMJCR1676],
the Japan Society for the Promotion of Science (JSPS)
(via the KAKENHI Grant No.~JP20H00134 and the
JSPS–RFBR Grant No.~JPJSBP120194828),
the Army Research Office (ARO) (Grant No.~W911NF-18-1-0358),
the Asian Office of Aerospace Research and Development (AOARD) (Grant No.~FA2386-20-1-4069), and
the Foundational Questions Institute Fund (FQXi) (Grant No.~FQXi-IAF19-06).
\end{acknowledgments}

\bibliography{Manuscript}

\end{document}


\title{ SUPPLEMENTAL MATERIAL:\\Multipartite Entanglement of the Topologically Ordered State in a Perturbed Toric Code}
\author{Yu-Ran Zhang}
\affiliation{Theoretical Quantum Physics Laboratory, RIKEN Cluster for Pioneering Research,
Wako-shi, Saitama 351-0198, Japan}

\author{Yu Zeng}
\affiliation{Institute of Physics, Chinese Academy of Sciences, Beijing 100190, China}
\author{Tao Liu}
\affiliation{School of Physics and Optoelectronics, South China University of Technology, Guangzhou 510640, China}
\affiliation{Theoretical Quantum Physics Laboratory, RIKEN Cluster for Pioneering Research,
Wako-shi, Saitama 351-0198, Japan}

\author{Heng Fan}
\email{hfan@iphy.ac.cn}
\affiliation{Institute of Physics, Chinese Academy of Sciences, Beijing 100190, China}
\affiliation{CAS Center for Excellence in Topological Quantum Computation, UCAS, Beijing 100190, China}

\author{J. Q. You}
\email{jqyou@zju.edu.cn}
\affiliation{Department of Physics, Zhejiang University, Hangzhou 310027, China}

\author{Franco Nori}
\email{fnori@riken.jp}
\affiliation{Theoretical Quantum Physics Laboratory, RIKEN Cluster for Pioneering Research, Wako-shi,
Saitama 351-0198, Japan}
\affiliation{RIKEN Center for Quantum Computing (RQC),   Wako-shi, Saitama 351-0198, Japan}
\affiliation{Physics Department, University of Michigan, Ann Arbor, Michigan 48109-1040, USA}

\date{\today}

\maketitle

\clearpage

\renewcommand{\theequation}{S\arabic{equation}}
\setcounter{equation}{0}  

\renewcommand{\thefigure}{S\arabic{figure}}
\setcounter{figure}{0}  

\renewcommand{\thetable}{S\arabic{table}}
\setcounter{table}{0}

\section{Transforming the toric code model to uncoupled transverse Ising chains via dual transformations}
The Hamiltonian of the toric code model with external fields can be divided into two mutually commutative parts
$\hat{H}_{\textrm{tc}}^{\textrm{field}}\equiv \hat{H}_{\textrm{tc}}+\hat{V}=\hat{H}^A+\hat{H}^B$ with
$[\hat{H}^A,\hat{H}^B]=0$,
where
\begin{align}
  \hat{H}^A&=-J^A\sum_s\hat{A}_s-\lambda_z\sum_{i\in h}\hat{\sigma}_i^z,\\
\hat{H}^B&=-J^B\sum_p\hat{B}_p-\lambda_x\sum_{i\in v}\hat{\sigma}_i^x.
\end{align}
For clarity, $(i,j)$ denotes the site of the original lattice and the dual lattice at row $i$
and column $j$; $b_i^{\langle j,j+1\rangle}$ denotes the bond connecting two sites $(i,j)$ and
$(i,j+1)$. For an odd row $i=2k-1$, $(2k-1,j)$ belongs to the original lattice,
while for an even row $i=2k$, $(2k-1,j)$ is located on the dual lattice [see Fig.~{\color{blue}1}(b) in the main text].
By introducing the effective spins with Pauli operators  via the dual transformations,
\begin{equation}
\hat{\tau}_{2k-1,j}^x
=\prod_{l\leq j}\hat{A}_{{2k-1},l},\hspace{0.2 in}
\hat{\tau}_{2k-1,j}^z=\hat{\sigma}_{b_{2k-1}^{\langle j,j+1\rangle}}^z,
\end{equation}
on the lattice site $(2{k-1},j)$, $\hat{H}^A$ becomes \cite{Zeng2016}
\begin{equation}
  \hat{H}^A=\sum_{k=1}^N \hat{{H}}_{2k-1},
  \end{equation}
  with
\begin{equation}
  \hat{{H}}_{2k-1}=-\sum_{j=1}^N(J^A\hat{\tau}^x_{{2k-1},j}\hat{\tau}^x_{{2k-1},{j+1}}+\lambda^z_j\hat{\tau}^z_{{2k-1},j}).
\end{equation}
Similarly, we can introduce effective spins on the dual lattice with Pauli operators via the dual transformations:
\begin{equation}
\hat{\tau}_{2k,j}^x
=\prod_{l\leq j}\hat{B}_{{2k},l},\hspace{0.2 in}
\hat{\tau}_{2k,j}^z=\hat{\sigma}_{b_{2k}^{\langle j,j+1\rangle}}^x,
\end{equation}
with which $\hat{H}^B$ is expressed as
 \begin{equation}
   \hat{H}^B=\sum_{k=1}^N \hat{{H}}_{2k},
   \end{equation}
where
\begin{equation}
  \hat{{H}}_{2k}=-\sum_{j=1}^N(J^B\hat{\tau}^x_{{2k},j}\hat{\tau}^x_{{2k},{j+1}}+\lambda^x_j\hat{\tau}^z_{{2k},j}).
\end{equation}


\section{Wilson loop in the toric code model after a quantum quench}
In the previous section, we show that using the dual transformations, the Hamiltonian of the toric code model
 with external fields can be expressed as uncoupled transverse-field Ising chains on different lines.
 Therefore, the out-of-equilibrium dynamics of the toric code model can be investigated using the
 results of the  transverse-field  Ising chain.

We consider a quantum quench of a transverse-field Ising chain at its ground state with a Hamiltonian:
\begin{equation}
\hat{H}_{\textrm{Ising}}=-\sum_{j=1}^N[\hat{\tau}_j^x\hat{\tau}_{j+1}^x+\lambda(t)\hat{\tau}_j^z],
\end{equation}
where the transverse field is changed from $\lambda_0$ to $\lambda$ abruptly at time $t=0$
\begin{equation}
\lambda(t)=\left\{\begin{array}{c c}\lambda_0,&\textrm{ for } t\leq0\\\lambda,&\textrm{ for }t>0\end{array}\right..
\end{equation}
It can be rewritten in terms of the Jordan-Wigner transformation $\hat{\tau}_j^z=(2\hat{c}_j^\dag \hat{c}_j-1)$,
$\hat{\tau}_j^+=\hat{c}_j^\dag\prod_{l=1}^{j-1}(1-2\hat{c}_l^\dag \hat{c}_l)$, and $\hat{\tau}_1^+=\hat{c}_1^\dag$ as
\begin{equation}
\hat{H}_{\textrm{Ising}}=\sum_j[(\hat{c}_j-\hat{c}_j^\dag)(\hat{c}_{j+1}^\dag+\hat{c}_{j+1})-\lambda (\hat{c}_j^\dag \hat{c}_j-\hat{c}_j \hat{c}_j^\dag)].
\end{equation}
In the thermodynamic limit $N\gg1$, we use the Fourier transformation (with $q$ being the wave vector),
\begin{equation}
  \hat{c}_j=\sum_q e^{-iqj}\hat{c}_q/\sqrt{N},
\end{equation}
to obtain the Bogoliubov-de Gennes Hamiltonian of a transverse-field Ising chain,
\begin{equation}
\hat{H}_{\textrm{Ising}}
=\sum_q\mathbb{C}_q^\dag\mathcal{H}_q\mathbb{C}_q,\end{equation} 
with $\mathbb{C}_q^\dag\equiv(\hat{c}_q^\dag,\hat{c}_{-q})$,
$y_q\equiv-\sin q$, $z_q\equiv-\lambda-\cos q$,
$\tan\Theta_q\equiv y_q/z_q$, and
\begin{equation}
{\mathcal{H}_q}=\left(\begin{array}{c c}-\lambda-\cos q & i\sin q\\-i\sin q & \lambda+\cos q\end{array}\right)
= y_q\hat{\sigma}^y+z_q\hat{\sigma}^z.
\end{equation}
The Hamiltonian can
be diagonalized by using the Bogoliubov transformation as
\begin{equation}
  \hat{H}_{\textrm{Ising}}=\sum_q\omega_q\mathbb{E}_q^\dag\hat{\sigma}^z\mathbb{E}_q,
  \end{equation}
with $\mathbb{C}_q=\mathbb{R}_q\mathbb{E}_q$, $\omega_q=(y_q^2+z_q^2)^{\frac{1}{2}}$,
$\mathbb{E}_q^\dag\equiv(\hat{\eta}_q^\dag,\hat{\eta}_{-q})$, 
\begin{equation}
\mathbb{R}_q=\left(\begin{array}{c c}u_q & -iv_q\\-iv_q & u_q\end{array}\right),
\end{equation}
$u_q\equiv\cos\frac{\Theta_q}{2}$, and $v_q\equiv\sin\frac{\Theta_q}{2}$.

The Heisenberg equation of $\mathbb{C}_q(t)$ 
can be expressed as
\begin{equation}\label{he}
i\partial_t\mathbb{C}_q(t)=2{\mathcal{H}_q} \mathbb{C}_q(t).
\end{equation}
We then define the time-propagation transformation of $\mathbb{C}_q(t)$ in terms of $\mathbb{E}_q(0)$  as $\mathbb{C}_q(t)=\mathbb{S}_q(t)\mathbb{E}_q(0)$,
where
\begin{equation}
\mathbb{S}_q(t)=\left(\begin{array}{c c}U_q(t) & -V_q^*(t)\\V_q(t) & U^*_q(t)\end{array}\right).
\end{equation}
Substituting it into the Heisenberg equation (\ref{he}), we can obtain
\begin{equation}\label{heq}
i\partial_t\mathbb{S}_q(t)=2\mathcal{H}_q(g)\mathbb{S}_q(t).
\end{equation}
With the initial constraints $U_q(0)=u_q(0)\equiv u_0$ and $V_q(0)=-iv_q(0)\equiv -iv_0$, 
the solutions of Eq.~(\ref{heq}) can be obtained as
\begin{align}
\left(\begin{array}{c}U_q(t)\\V_q(t)\end{array}\right)
=\left(\begin{array}{c}u_0\cos2\omega t+\frac{i\sin2\omega t(zu_0+yv_0)}{\omega}\\
-iv_0\cos2\omega t+\frac{\sin2\omega t(yu_0+zv_0)}{\omega}\end{array}\right),
\end{align}
where $\omega\equiv\omega_q(g)$, $y\equiv y_q(g)$, and $z\equiv z_q(g)$.

To calculate the reduced Wilson loop,  written as an $x$-directional spin correlation function [Eq.~({\color{blue}5}) in the main text], we need to consider two kinds of operators
$\hat{\mathcal{A}}_j(t)=\hat{a}_j^\dag(t)+\hat{a}_j(t)$ and $\hat{\mathcal{B}}_j(t)=\hat{b}_j^\dag(t)-\hat{b}_j(t)$, where
\begin{align}
\hat{a}_j(t)&=\sum_{q}e^{iqj}[U_q(t)+V_q(t)]\hat{\eta}_q(0)/\sqrt{N},\\
\hat{b}_j(t)&=\sum_{q}e^{iqj}[U_q(t)-V_q(t)]\hat{\eta}_q(0)/\sqrt{N}.
\end{align}
According to Wick's theorem, we only need to consider three types of contraction \cite{Zeng2016},
\begin{align}
&G_{r}(t)\equiv\langle\mathcal{G}(0)|\hat{\mathcal{B}}_j(t)\hat{\mathcal{A}}_{j+r}(t)|\mathcal{G}(0)\rangle\nonumber\\
&=\sum_q \frac{e^{-iqr}}{N}\left[\frac{z_0z+y_0y}{(iy-z)\omega_0}+i\frac{z_0y-zy_0}{(iy-z)\omega_0}\cos4\omega t\right],\\
&G_{r}^A(t)\equiv\langle\mathcal{G}(0)|\hat{\mathcal{A}}_j(t)\hat{\mathcal{A}}_{j+r}(t)|\mathcal{G}(0)\rangle\nonumber\\
&=\delta_{r,0}+\sum_q \frac{e^{-iqr}}{N}\frac{z_0y-zy_0}{\omega\omega_0}\sin4\omega t,\\
&G_{r}^B(t)\equiv\langle\mathcal{G}(0)|\hat{\mathcal{B}}_j(t)\hat{\mathcal{B}}_{j+r}(t)|\mathcal{G}(0)\rangle\nonumber\\
&=-\delta_{r,0}+\sum_q \frac{e^{-iqr}}{N}\frac{z_0y-zy_0}{\omega\omega_0}\sin4\omega t.
\end{align}
The $x$-directional spin correlation function, expressed as a Pfaffian, can be calculated.
Details of the calculation can be found in Refs.~\cite{BAROUCH1971,Suzuki2012}.

We consider the quench from the ground state of the toric code $\lambda_0=0$ to the case with $\lambda_f=\lambda$.
In the long-time limit ($t\rightarrow\infty$) and thermodynamic limit ($\frac{1}{N}\sum_q\rightarrow\frac{1}{2\pi}\int\! dq$),
the time-dependent oscillating terms vanish, and therefore, we can obtain that $G_{r}^A=\delta_{r,0}$, $G_{r}^B=-\delta_{r,0}$,  for $\lambda<1$
\begin{align}
G_r(\infty)=\left\{\begin{array}{ll}(1-\lambda^2)\lambda^{r-1}/2,&\textrm{ for }r\geq2\\
1-{\lambda^2}/{2},&\textrm{ for }r=1\\
-{\lambda}/{2},&\textrm{ for }r=0\\
0,&\textrm{ for }r\leq-1
\end{array}\right.,
\end{align}
and for $\lambda>1$
\begin{align}
G_r(\infty)=\left\{\begin{array}{ll}0,&\textrm{ for }r\geq2\\
1/2,&\textrm{ for }r=1\\
-1/(2\lambda),&\textrm{ for }r=0\\
(\lambda^2-1)\lambda^{r-1}/2,&\textrm{ for }r\leq-1
\end{array}\right.,
\end{align}
by using the residue theorem.
Furthermore, the $x$-directional spin correlation function $C_d^x(\infty)\equiv\langle\sigma_j^x(\infty)\sigma_{j+d}^x(\infty)\rangle_{\mathcal{G}}$ (without loss of generality, we assume that $d\geq0$) reduces to a determinant
\begin{equation}
C_d^x(\infty)=\left|\begin{array}{cccc}G_1(\infty)&G_0(\infty)&\cdots&G_{-d+2}(\infty)\\
G_2(\infty)&G_1(\infty)&\cdots&G_{-d+3}(\infty)\\
\vdots&\vdots&\ddots&\vdots\\
G_{d}(\infty)&G_{d-1}(\infty)&\cdots&G_{1}(\infty)\end{array}\right|.
\end{equation}
For $\lambda>1$, we can obtain that $C^x_d(\infty)=1/2^d$,
which decays exponentially with respect to the distance $d$.
For $0<\lambda<1$, the result becomes \cite{Sengupta2004}
\begin{align}
C^x_d(\infty)=\frac{\lambda^{d+1}}{2^d}\cosh\left[(d+1)\log\frac{1+\sqrt{1-\lambda^2}}{\lambda}\right],
\end{align}
which for $d\rightarrow\infty$ becomes
\begin{equation}
C^x_d(\infty)\rightarrow\left(\frac{1+\sqrt{1-\lambda^2}}{2}\right)^{d+1}.
\end{equation}

\section{Numerical evidence for Dynamical localization of topological order in a disordered toric code model}
Here, we consider a disordered Ising chain with the Hamiltonian:
\begin{equation}
\hat{H}^{\textrm{D/O}}=-\sum_j(J_j\hat{\tau}_j^x\hat{\tau}_{j+1}^x+\lambda_j\hat{\tau}_j^z).
\end{equation}
It can be rewritten in terms of the Jordan-Wigner transformation $\hat{\tau}_j^z=(2\hat{c}_j^\dag \hat{c}_j-1)$,
$\hat{\tau}_j^+=\hat{c}_j^\dag\prod_{l=1}^{j-1}(1-2\hat{c}_l^\dag \hat{c}_l)$, and $\hat{\tau}_1^+=\hat{c}_1^\dag$ as
\begin{align}
\hat{H}^{\textrm{D/O}}&=\sum_j[J_j(\hat{c}_j-\hat{c}_j^\dag)(\hat{c}_{j+1}^\dag+\hat{c}_{j+1})-\lambda_j(\hat{c}_j^\dag \hat{c}_j-\hat{c}_j \hat{c}_j^\dag)]\nonumber\\
&=\frac{1}{2}\mathbb{C}^\dag\mathbb{M}\mathbb{C},
\end{align}
where $\mathbb{C}^\dag=(\hat{c}_1^\dag,\cdots,\hat{c}_N^\dag,\hat{c}_1,\cdots,\hat{c}_N)$, and $\mathbb{M}=(^{\mathbb{A}}_{\mathbb{B}^T}{}_{-\mathbb{A}}^{{~\mathbb{B}}})$, with $\mathbb{A}$ and
$\mathbb{B}$ being $N\times N$ matrices of elements: $\mathbb{A}_{j,j}=2\lambda_j$, $\mathbb{A}_{j,j+1}=\mathbb{A}_{j+1,j}=-J_j$, $\mathbb{B}_{j,j+1}=-\mathbb{B}_{j+1,j}=-J_j$, and  boundary
conditions $\mathbb{A}_{N,1}=\mathbb{A}_{1,N}=J_N$, $\mathbb{B}_{N,1}=-\mathbb{B}_{1,N}=J_N$ for even parity.
By diagonalizing the matrix $\mathbb{R}\mathbb{M}\mathbb{R}^T=\mathbb{V}$,
the Hamiltonian can be diagonalized as
\begin{equation}
  \hat{H}^{\textrm{D/O}}=\frac{1}{2}\mathbb{E}^\dag\mathbb{V}\mathbb{E}=\sum_j\omega_j(\hat{\eta}_j^\dag\hat{\eta}_j-1/2),
\end{equation}
where $\mathbb{C}=\mathbb{R}^T\mathbb{E}$ with the orthogonal matrix  $\mathbb{R}=(^{\mathbb{G}}_{\mathbb{H}}{~}_{\mathbb{G}}^{{\mathbb{H}}})$, $\mathbb{E}^\dag=(\hat{\eta}_1^\dag,\cdots,\hat{\eta}_L^\dag,\hat{\eta}_1,\cdots,\hat{\eta}_L)$,
and $\mathbb{V}=(^{\mathbb{W}}{~}_{-\mathbb{W}})$, with the diagonal matrix
 $\mathbb{W}=\textrm{diag}(\omega_1,\cdots,\omega_N)$.

 Similarly, the Heisenberg equation of $\hat{\eta}_j(t)$ is expressed as
 \begin{equation}
   i\partial_t\hat{\eta}_j(t)=i[\hat{H}^{\textrm{D/O}},\hat{\eta}_j(t)]=-i\omega_j\hat{\eta}_j,
 \end{equation}
 of which the solution can be expressed as
 \begin{equation}
   \mathbb{E}(t)=\left(
   \begin{array}{cc}
     e^{-i\mathbb{W}t}&\\&e^{i\mathbb{W}t}
     \end{array}\right)\mathbb{E}
     =\exp(-i\mathbb{V}t)\mathbb{E}.
 \end{equation}
Considering that an initial Hamiltonian $\hat{H}_{0}^{\textrm{D/O}}$ is quenched to the new Hamiltonian $\hat{H}^{\textrm{D/O}}$,
 we have
 \begin{equation}\label{expa}
   \mathbb{C}(t)=\mathbb{R}^T\exp(-i\mathbb{V}t)\mathbb{R}\mathbb{C}_0,
 \end{equation}
 where $\mathbb{R}$ and $\mathbb{V}$ are for the quenched Hamiltonian  $\hat{H}^{\textrm{D/O}}$, $\mathbb{C}_0$ is for the initial Hamiltonian $\hat{H}_{0}^{\textrm{D/O}}$, and $\langle\cdots\rangle_{\mathcal{G}}$ denotes the average with the ground state of the initial Hamiltonian
 $\hat{H}_{0}^{\textrm{D/O}}$.

Then, the $x$-directional spin correlation function $C_d^x(t)\equiv\langle\sigma_j^x(t)\sigma_{j+d}^x(t)\rangle_{\mathcal{G}}$ (without loss of generality, we assume that $d\geq0$) can be expressed as
\begin{equation}
  C_d^x(t)=\langle \hat{\mathcal{B}}_j(t)\hat{\mathcal{A}}_{j+1}(t)\hat{\mathcal{B}}_{j+1}(t)\cdots \hat{\mathcal{B}}_{j+d-1}(t)\hat{\mathcal{A}}_{j+d}(t)\rangle_{\mathcal{G}},
\end{equation}
where $\hat{\mathcal{A}}_j\equiv \hat{c}_j^\dag+\hat{c}_j$ and $\hat{\mathcal{B}}_j\equiv \hat{c}_j^\dag-\hat{c}_j$.
Using Wick's theorem, the spin correlation function can be written as a Pfaffian of
a skew-symmetric matrix $\mathbb{T}$
\begin{equation}
  C_d^x(t)=\textrm{pf}[\mathbb{T}(j,j+d,t)].\label{pfa}
\end{equation}
The skew-symmetric matrix $\mathbb{T}(i,j,t)$ is defined as
\begin{equation}
\mathbb{T}(i,j,t)\equiv\left(
\begin{array}{cc}
  \mathbb{P}(i,j,t)&\mathbb{M}(i,j,t)\\
  -\mathbb{M}(i,j,t)^T&\mathbb{Q}(i,j,t)
  \end{array}
  \right),\label{matrix}
\end{equation}
of which the elements of the submatrices are
\begin{align}
\mathbb{P}_{mn}(i,j,t)&=\delta_{mn}+\langle \hat{\mathcal{B}}_{j+m-1}(t)\hat{\mathcal{B}}_{j+n-1}(t)\rangle_{\mathcal{G}},\label{e1}\\
\mathbb{Q}_{mn}(i,j,t)&=-\delta_{mn}+\langle \hat{\mathcal{A}}_{j+m}(t)\hat{\mathcal{A}}_{j+n}(t)\rangle_{\mathcal{G}},\\
\mathbb{M}_{mn}(i,j,t)&=\langle \hat{\mathcal{B}}_{j+m-1}(t)\hat{\mathcal{A}}_{j+n}(t)\rangle_{\mathcal{G}}.\label{e3}
\end{align}
By expanding $\hat{\mathcal{A}}_j(t)$ and $\hat{\mathcal{B}}_j(t)$ with Eq.~(\ref{expa}), we have
\begin{align}
  \langle \hat{\mathcal{A}}_{m}(t)\hat{\mathcal{A}}_{n}(t)\rangle_{\mathcal{G}}&=[\tilde{\Phi}(t)\tilde{\Phi}(t)^\dag]_{mn},\label{i1}\\
\langle \hat{\mathcal{B}}_{m}(t)\hat{\mathcal{B}}_{n}(t)\rangle_{\mathcal{G}}&=-[\tilde{\Psi}(t)\tilde{\Psi}(t)^\dag]_{mn},\\
\langle \hat{\mathcal{A}}_{m}(t)\hat{\mathcal{B}}_{n}(t)\rangle_{\mathcal{G}}&=[\tilde{\Phi}(t)\tilde{\Psi}(t)^\dag]_{mn},\\
\langle \hat{\mathcal{B}}_{m}(t)\hat{\mathcal{A}}_{n}(t)\rangle_{\mathcal{G}}&=-[\tilde{\Psi}(t)\tilde{\Phi}(t)^\dag]_{mn},\label{i4}
\end{align}
with
\begin{align}
  \tilde{\Phi}(t)&=\Phi^T\cos(\mathbb{W}t)\Phi\Phi_0^T-i\Phi^T\sin(\mathbb{W}t)\Psi\Psi_0^T,\label{d1}\\
\tilde{\Psi}(t)&=\Psi^T\cos(\mathbb{W}t)\Psi\Psi_0^T-i\Psi^T\sin(\mathbb{W}t)\Phi\Phi_0^T,\label{d2}
\end{align}
where $\Phi=\mathbb{G}+\mathbb{H}$ and $\Psi=\mathbb{G}-\mathbb{H}$ are for the quenched Hamiltonian $\hat{H}^{\textrm{D/O}}$; $\Phi_0=\mathbb{G}_0+\mathbb{H}_0$ and $\Psi_0=\mathbb{G}_0-\mathbb{H}_0$ are for the initial Hamiltonian $\hat{H}_{0}^{\textrm{D/O}}$.
Therefore, the evolution of the reduced Wilson loop [Eq.~({\color{blue}5}) in the main text] can be numerically simulated by calculating the Pfaffian (\ref{pfa}) of a matrix (\ref{matrix}) with elements (\ref{e1}--\ref{e3}) using (\ref{i1}--\ref{i4})
with equations of evolutions (\ref{d1},\ref{d2}) (see also Ref.~\cite{Zeng2021}).


\section{Upper bound of the quantum Fisher information of the thermal state of the toric code model }
The thermal state of the toric code model at temperature $T$  with a Hamiltonian in Eq.~({\color{blue}1}) in the main text
can be factorized as
\begin{align}
&\rho_T=\frac{e^{-\hat{H}_{\textrm{tc}}/T}}{\mathcal{Z}}\\
&\sim{\prod_{s}[1+\tanh(J^A/T)\hat{A}_s]\prod_{p}[1+\tanh( J^B/T)\hat{B}_p]},
\end{align}
with the partition function being
$\mathcal{Z}=\textrm{Tr}[\exp(-\hat{H}_{\textrm{tc}}/T)]$.

To obtain the upper bound of the quantum Fisher information  [Eq.~({\color{blue}11}) in the main text] with respect
to the generator $\hat{\mathcal{O}}^{\textrm{m}}$ [Eq.~({\color{blue}6}) in the main text],
we first choose the   state $|l\rangle$ as a product state
in the $\hat{\sigma}^z$ basis, and the minimally entangled typical thermal state (METTS) is expressed as
\begin{equation}
    |\phi_l\rangle=\rho_T^{1/2}|l\rangle/\sqrt{p_l}
    \sim\prod_{s}\{1+\tanh[J^A/(2T)]\hat{A}_s\}|l\rangle,
\end{equation}
with $p_l=\langle l|\rho_T|l\rangle$.
Then, we calculate the average variance of the
generator $\hat{\mathcal{O}}^{\textrm{m}}$ [Eq.~({\color{blue}6}) in the main text] for the METTSs
$\{|\phi_l\rangle\}$ as
\begin{align}
&F_Q[\hat{\mathcal{O}}^{\textrm{m}},\rho_T]\leq\sum_lp_l(\Delta_{\phi_l} \hat{\mathcal{O}}^{\textrm{m}})^2\\
&=L^2\sum_lp_l\left(1+\sum_{D=1}^{L-1}\langle\phi_l|\hat{\tau}_1^x\hat{\tau}_{1+D}^x|\phi_l\rangle\right)\nonumber\\
&=L^2\left(1+\sum_{D=1}^{L-1}\sum_lp_l\langle\phi_l|\hat{\tau}_1^x\hat{\tau}_{1+D}^x|\phi_l\rangle\right)\nonumber\\
&=L^2\left(1+\sum_{D=1}^{L-1}\sum_l\langle l|\rho^{1/2}\hat{\tau}_1^x\hat{\tau}_{1+D}^x\rho^{1/2}|l\rangle\right)\nonumber\\
&=L^2\left[1+\sum_{D=1}^{L-1}\langle l|\hat{\tau}_1^x\hat{\tau}_{1+D}^x\sum_{t=0,1}\prod_i[\tanh (J^A/T)\hat{\tau}_i^x\hat{\tau}_{i+1}^x]^t|l\rangle\right]\nonumber\\
&=L^2\left[1+\sum_{D=1}^{L-1}\tanh (J^A/T)^D\right],
\end{align}
where the first index $(2k-1)$ of the dual Pauli operators $\hat{\tau}_{2k-1,j}$ for the original lattice have been omitted,
and we have used the fact that $\langle l|\hat{\tau}_{2k-1,j}^x|l\rangle=0$ and $\langle\phi_l|\hat{\mathcal{O}}^{\textrm{m}}|\phi_l\rangle=0$.

Similarly, we consider the quantum Fisher information with respect to the generator $\hat{\mathcal{O}}^{\textrm{e}}$.
We choose the state $|l'\rangle$ as a product state in the $\hat{\sigma}^x$ basis, and the METTS is written as
\begin{equation}
    |\phi_{l'}\rangle=\rho_T^{1/2}|l'\rangle/\sqrt{p_{l'}}
    \sim\prod_{p}\{1+\tanh[ J^B/(2T)]\hat{B}_p\}|l'\rangle,
\end{equation}
with $p_{l'}=\langle l'|\rho_T|l'\rangle$. The upper bound of the quantum Fisher information with respect to the
generator $\hat{\mathcal{O}}^{\textrm{m}}$ [Eq.~({\color{blue}6}) in the main text] for the METTSs
$\{|\phi_{l'}\rangle\}$ can be calculated as
\begin{align}
F_Q[\hat{\mathcal{O}}^{\textrm{e}},\rho_T]&\leq\sum_{l'}p_{l'}(\Delta_{\phi_{l'}} \hat{\mathcal{O}}^{\textrm{e}})^2\\
&=L^2\left[1+\sum_{D=1}^{L-1}\tanh (J^B/T)^D\right].
\end{align}


\bibliography{supp}